\documentclass[10pt,preprint]{aastex} 
 
\usepackage{amsmath}
\usepackage{amssymb}
\usepackage{color}
\begin{document}

\title{Measuring Mass Accretion Rate onto the Supermassive Black Hole in M 87 Using Faraday Rotation Measure with the Submillimeter Array}

\author{C. Y. Kuo\altaffilmark{1},  K. Asada\altaffilmark{1}, R. Rao\altaffilmark{1}, M. Nakamura\altaffilmark{1}, J. C. Algaba\altaffilmark{1},  
H. B. Liu\altaffilmark{1}, M. Inoue\altaffilmark{1}, P. M. Koch\altaffilmark{1}, P. T. P. Ho\altaffilmark{1}, S. Matsushita\altaffilmark{1},
H.-Y. Pu\altaffilmark{1}, K. Akiyama\altaffilmark{2,3}, H. Nishioka\altaffilmark{1}, N. Pradel\altaffilmark{1} }
 
\affil{\altaffilmark{1}  Academia Sinica, Institute of Astronomy and Astrophysics, PO Box 23-141, Taipei 10617, Taiwan, R.O.C.}                                              
\affil{\altaffilmark{2} National Astronomical Observatory of Japan, Osawa 2-21-1, Mitaka, Tokyo 181-8588, Japan}                                
\affil{\altaffilmark{3} Department of Astronomy, Graduate School of Science, The University of Tokyo, 7-3-1 Hongo, Bunkyo-ku, Tokyo 113-0033, Japan }

\begin{abstract} 
We present the first constraint on Faraday rotation measure (RM) at submillimeter wavelengths for the nucleus of M 87. By fitting the polarization position angles ($\chi$) observed with the SMA at four independent frequencies 
around $\sim$230 GHz and interpreting the change in $\chi$ as a result of \emph{external} Faraday rotation associated with accretion flow, we determine the rotation measure of the M 87 core to be between $-$7.5$\times$10$^{5}$ and 3.4$\times$10$^{5}$ rad/m$^{2}$. Assuming a density profile of the accretion flow that follows a power-law distribution and a magnetic field that is ordered, radial, and has equipartition strength, the limit on the rotation measure constrains the mass accretion rate $\dot{M}$ to be below 9.2$\times$10$^{-4}$ M$_{\odot}$~yr$^{-1}$ at a distance of 21 Schwarzchild radii from the central black hole. This value is at least two orders of magnitude smaller than the Bondi accretion rate, suggesting significant suppression of the accretion rate in the inner region of the accretion flow. Consequently, our result disfavors the classical \emph{advection dominated accretion flow} (ADAF) and prefers the \emph{adiabatic inflow-outflow solution} (ADIOS) or \emph{convection-dominated accretion flow} (CDAF) for the hot accretion flow in M 87. 

\end{abstract} 
 

\section{INTRODUCTION}

Revealing the mass accretion process onto supermassive black holes
(SMBHs) is crucial for understanding the nature of AGNs because it is
believed that the mass accretion fuels a SMBH, supports its nuclear
luminosity, and presumably powers its outflow. Given the simplest
approximation of an axisymmetric adiabatic accretion flow onto a SMBH,
Bondi (1952) derived a black hole accretion power $P_{\rm
B}=\dot{M}_{\rm B} c^2$, where $\dot{M}_{\rm B}$ is the Bondi accretion
rate which is defined at the sphere of gravitational influence of the SMBH (i.e., the
Bondi radius $r_{\rm B}$).  

The active galaxy M 87, one of the well-known LLAGNs, possesses a huge
black hole mass $M_{\bullet}=(3.2$--$6.6) \times10^{9} M_{\odot}$
(Macchetto et al. 1997; Gebhardt et al. 2011; Walsh et al. 2013) with a
prominent relativistic jet\footnote{We adopt 6.6 $\times$ 10$^{9} M_{\odot}$ for the black hole mass in this paper.}.  
The estimated jet power is in a range between $\sim10^{42}$ to $10^{44}$ erg s$^{-1}$ (Li et al. 2009 and
references therein) and seems to match fairly well with the overall energetics described by the
Bondi accretion\footnote{$\dot{M}_{\rm B} \simeq 0.12 M_{\odot}\,$yr$^{-1}$ is adopted here.}
($P_{\rm B} \sim 7 \times 10^{45}$ erg s$^{-1}$) at $r_{\rm B}\simeq
230$ pc. However, the
central core of M 87 is highly under-luminous in the X-ray band and its X-ray luminosity is smaller than $P_{\rm B}$ by
five orders of magnitude ($L_{\rm X} \sim 7 \times 10^{40}$ erg~s$^{-1}$; Di Matteo et al. 2003). The low X-ray
luminosity implies the existence of radiatively inefficient accretion flows (RIAFs)
and/or a substantial decrease of the mass accretion rate
$\dot{M}  \ll \dot{M_{\rm B}}$ at radii $r \ll r_{\rm
B}$. 

The theory of RIAFs, such as the advection-dominated accretion flow
(ADAF: Ichimaru 1977; Narayan \& Yi 1995), convection-dominated
accretion flow (CDAF: Narayan 2000; Quataert \& Gruzinov 2000), and
adiabatic inflow-outflow solution (ADIOS: Blandford \& Begelman 1999) was developed over the last decades. The resulting mass accretion
rates toward an SMBH normalized by the Bondi accretion rate can be
scaled as a function of the spherical radius as
$\dot{M} /\dot{M}_{\rm B}=(r/r_{\rm B})^{\kappa}$ with $\kappa=0 -      
1$. Considering the viscosity parameter $\alpha \geq 0.01$
(Shakura \& Sunyaev 1973), solutions of classical ADAFs suggest a mass
accretion rate from $r_{\rm B}$ to $r_{\rm s}$ that is comparable to $\dot{M}_{\rm B}$ : $\dot{M}_{\rm GADAF} \sim (0.1 -
1)\times \dot{M}_{\rm B}$ (GADAF stands for ``giant''
ADAF; Narayan \& Fabian 2011). On the other hand, CDAFs exhibit a
substantial decrease of the mass accretion rate toward $r \simeq r_{\rm s}$
as $\dot{M}_{\rm CDAF} \sim (r/r_{\rm B}) \times \dot{M}_{\rm B}$ (Igumenshchev
\& Narayan 2002). The ADIOS, which generalizes the ADAF model by
including the disk wind, can take intermediate values of $\kappa$
anywhere between 0 (ADAF) and 1 (CDAF). Recent
numerical simulations favor $\kappa \sim 0.4$--0.7 which is consistent 
with ADIOS (e.g., Pang et al. 2011;
Yuan et al. 2012).

There has been a growing consensus during the last decade that
millimeter (mm)/submillimeter (submm) polarimetry provides useful
diagnostics to infer $\dot{M} $ of RIAFs at $r \ll r_{\rm B}$
(Agol 2000; Quataert \& Gruzinov 2000). In particular, through 
the Faraday rotation measure (RM; an integral of the product of the thermal electron
density and the magnetic field component along the line of
sight) of the linear polarization, $\dot{M_{\bullet}}$ toward Sgr A* was
examined (e.g., Bower et al. 2003; Marrone et al. 2006; Macquart et
al. 2006) and an upper limit of $\dot{M}$ was found to be $\leq 10^{-7}$--$10^{-6}
M_{\odot}\,$yr$^{-1}$ at $r \lesssim 100\, r_{\rm s}$ ($r_{\rm s}$ is the Schwarzschild radius).  We note
that the submm wavelengths are more advantageous than other radio bands
for determining the RM because the bandwidth depolarization is less
significant, and the opacities of the accretion flow and the jet are smaller so that one can 
probe regions closer to the BH. However, efforts to determine the $\dot{M} $
of AGNs (including M 87) have never been undertaken except for Sgr A*.

For M 87, all previous efforts on measuring the RM primarily focused on
either the jet or radio lobes at centimeter wavelengths.  Owen et
al. (1990) found RM values of a few thousands rad/m$^{2}$ at lobes 
(interpreted as arising from a foreground Faraday screen) and of a
few hundreds rad/m$^{2}$ in the kiloparsec-scale jet.  More recently, Algaba et al. (2013) found similar RM
values with a hint of gradients across the kpc jet.  On upstream parsec
scales, Junor et al. (2001) found an average RM of $\sim$$-$4400
rad/m$^{2}$ with regions that show both positive and negative signs.
Zavala \& Taylor (2002) found similar results and explained the origin 
of the observed RM with foreground medium not directly associated with the jet.  
To our knowledge, no mm/submm polarization observations for examining RM toward M 87 have
been conducted.

In this letter, we present the first measurement of RM toward the
M 87 nucleus at submm wavelengths. We also derive the mass accretion rate $\dot{M}$ 
onto the black hole based on the measured RM and provide
a constraint on RIAF models. In section 2, we introduce our SMA observation
of the RM together with our data reduction. In section 3, we show our main results. 
The estimation of $\dot{M}$, the discussion on the preferred RIAF models, and alternative interpretations of RM are presented in section 4.

\section{Observations and Data Reduction}

M 87 was observed in the 230 GHz band on 2013 January 23 with the Submillimeter Array (SMA; Ho et al. 2004)\footnote{The Submillimeter Array
is a joint project between the Smithsonian Astrophysical Observatory
and the Academia Sinica Institute of Astronomy and Astrophysics and
is funded by the Smithsonian Institution and the Academia Sinica.}.
The observation was conducted with seven antennas in the extended array configuration. The weather condition was excellent during the observation with an atmospheric opacity $\tau_{225} \sim 0.05$.
The total length of time on M 87 was 8 hours. 
The SMA receivers operate in double sideband mode with each sideband having a width of 4 GHz. 
Each sideband was further split into a pair of 2-GHz-wide intermediate frequency (IF) sub-bands. 
We centered those IFs at 230.3 and 232.3 GHz in the upper sideband (USB) and at 218.4 and 220.4 GHz in the lower sideband (LSB), respectively. 
The SMA polarimeter was used to sample all four polarized correlations (LL, LR, RL, and RR) by switching polarization between the left-hand and right-hand circularly polarized feeds in period with 16 Walsh function patterns. 
A detailed discussion of the SMA polarimetry system is given in Marrone (2006) and Marrone \& Rao (2008).

We performed initial flagging and calibrations including the flux, bandpass, and gain calibrations in the MIR-IDL package developed for the SMA. We conducted polarization calibration, imaging, and data analysis in the MIRIAD package (Wright \& Sault 1993). The flux calibration was done with measurements of Callisto, and we performed bandpass, amplitude, and phase calibration with frequent observations of the quasar 3C279, which is 19$^{\circ}$.2 away from M 87. To perform polarization calibration, we observed 3C279 and 3C84 over a large range of parallactic angles and solved for quasar polarization and leakage terms. The average values of the instrumental polarization (D-terms)  or ``leakage'' were approximately 1$-$2\% in the upper sideband and 4$-$5\% in the lower sideband. The leakage terms derived from 3C279 and 3C84 were in excellent agreement with each other. We estimated the accuracy of the leakage terms to be $\sim$0.5\%. 

After deriving the  D-terms, we corrected the gain of M 87 data by self-calibration (phase-only) techniques in order to increase the dynamic range, and then applied polarization leakages. We performed imaging and deconvolution 
with I, Q, U, and V imaged individually for each IF band in each sideband, leading to four sets of I, Q, U, V images. The final synthesized beams for the continuum images are 1".2$\times$0".8 (corresponding to $\sim$ (1.7 $\times$ 1.0 ) $\times$10$^{5}$$r_{s}$). Finally, we used the task IMPOL in MIRIAD along with the Q and U maps to derive total linearly polarized intensity and distribution of the electric vector position angles (EVPA). The magnetic field distribution of M 87 were obtained by rotating the EVPAs by 90$\arcdeg$.
\section{The Rotation Measure of the M 87 Core} 
In Figure 1 we show a sample 
Stokes images of M 87 in the 232.3 GHz band. Figure 2 shows the magnetic field distribution. One can see that the magnetic vector orientations are perpendicular to the jet in knots A and C and parallel in knot B. This is in good agreement with polarization characteristics derived from observations at centimeter wavelengths and optical band (e.g., Owen et al. 1989, Perlman et al. 1999), indicating the robustness of our polarization observation and calibration. In table 1, we show the I, Q, U fluxes, polarization fractions, and the EVPA of the M 87 core for each of the four bands in the observation. In the following, we will interpret the change of the polarization position angle $\chi$ as a result of \emph{external} Faraday rotation associated with accretion flow. The reason why we prefer this interpretation will be explained in section 4.2.

Faraday rotation changes the observed $\chi$ as a function of observing wavelength $\lambda$ according to
\begin{equation*}
\chi(\lambda) = \chi_{0}+RM\lambda^{2}~,
\end{equation*}
where $\chi_{0}$ is the intrinsic polarization angle and RM is the rotation measure. We determine the RM of the M 87 core by fitting the polarization position angle ($\chi$ $=$ 0.5$\,\arctan$(U/Q)) from four
independent frequencies within the SMA band based on the Q and U maps from the observation. The Q and U values for RM are measured by taking the peak fluxes of the Q and U maps, with the image rms uncertainty as the measurement error for Q and U. The uncertainty in $\chi$ is calculated from an error propagation of the Q and U uncertainties.

\begin{figure}[ht] 
\begin{center} 
\vspace*{0.0 cm} 
\hspace*{0.0 cm} 
\includegraphics[angle=0, scale=0.3]{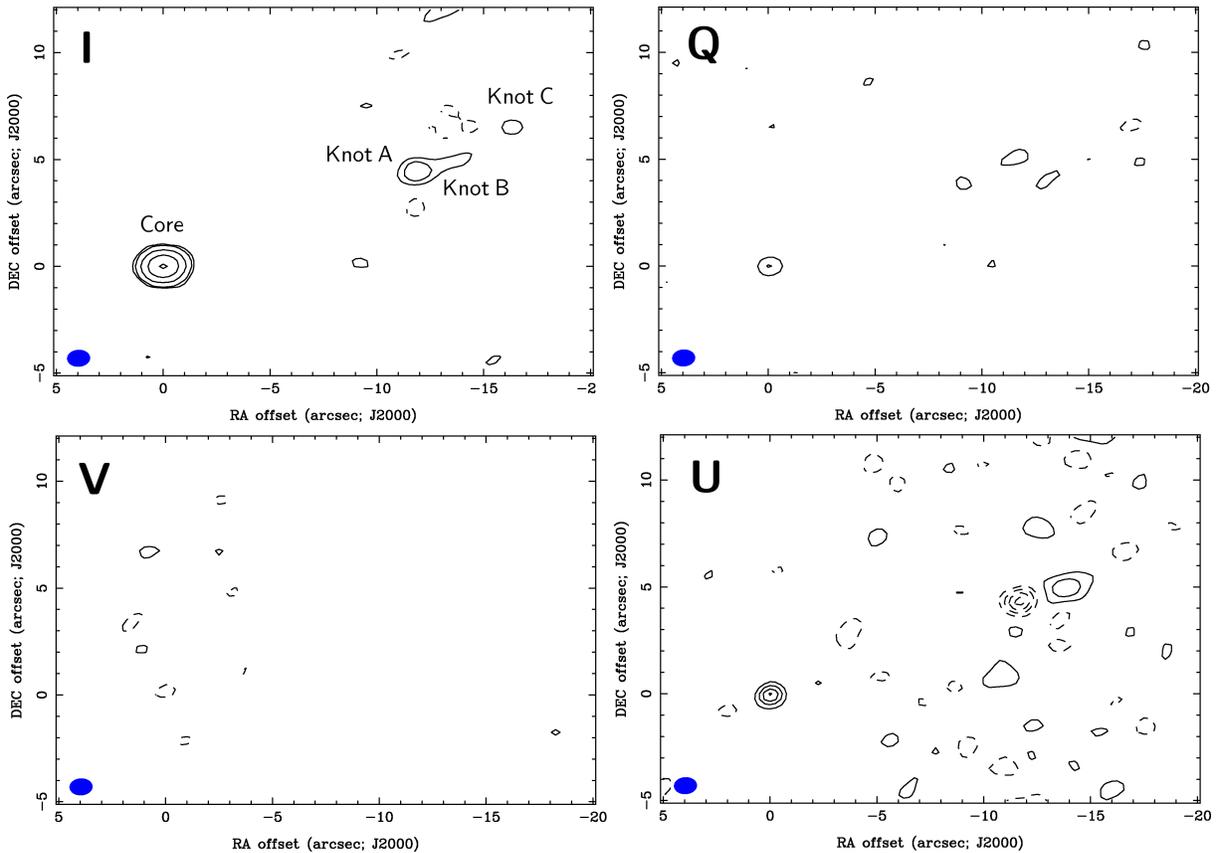}
\vspace*{0.0 cm} 
\caption{Sample Stokes images of M 87 from the IF2 band of the USB data. The synthesized beam  (bottom left in each panel) is 1".2$\times$0".8. Panels clockwise from top left show $I$, $Q$, ~$U$, and ~$V$. For Stokes $I$, we draw contours at $-$10, $-$5, 5, 10, 40, 160, and 480 times the 3 mJy~beam$^{-1}$ rms noise in the image.
For $Q$, ~$U$, and ~$V$, we draw contours at $-$12, $-$9, $-$6, $-$3, 3, 6, 9, and 12 times the rms noise in the image, which is 1.3, 1.3, and 2.2 mJy~beam$^{-1}$ for $Q$, ~$U$, and ~$V$, respectively.    }
\end{center} 
\end{figure}  

\begin{figure}[ht] 
\begin{center} 
\vspace*{0.0 cm} 
\hspace*{0.0 cm} 
\includegraphics[angle=0, scale=0.5]{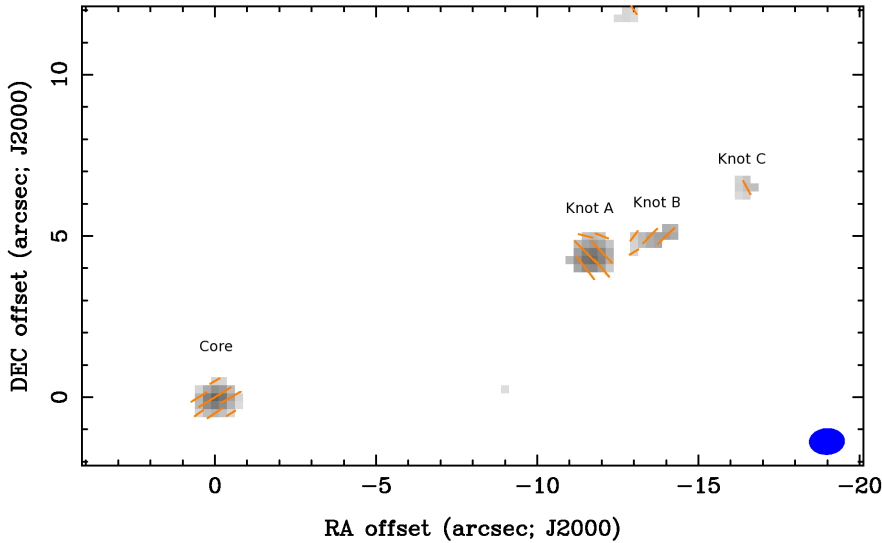}
\vspace*{0.0 cm} 
\caption{Total intensity of M 87 at the 232.3 GHz band in grey scale. The orange segments show the magnetic vector position angles (MVPA) derived for the same band. The MVPAs were obtained by rotating EVPAs by 90$^{\circ}$. The clumps from left to right in the figure are the nucleus (core), jet knot A, B, and C, respectively. The synthesized beam is shown in the bottom right of the image. }
\end{center} 
\end{figure} 

\begin{deluxetable}{lccclrl} 
\tablewidth{0 pt} 
\tablecaption{230 GHz Polarization Measurements of M 87 nucleus} 
\tablehead{ 
\colhead{Frequency}  & \colhead{$\nu$}     & \colhead{$I$}
 & \colhead{$Q$}  &                                                 
\colhead{$U$}   & \colhead{$m$} & \colhead{$\chi$}
\\                                                                      
\colhead{Band}  & \colhead{(GHz)} & \colhead{(Jy)}  &\colhead{(mJy)}
&                                                                       
\colhead{(mJy)}        & \colhead{(\%)}    & \colhead{(deg)}     } 
\startdata 
USB; IF 1  & 230.3 &  1.602$\pm$0.003    & 8.1$\pm$1.3   & 14.7$\pm$1.3   &  1.05$\pm$0.08 &  30.7$\pm$2.2 \\
USB; IF 2  & 232.3 &  1.594$\pm$0.003    & 8.2$\pm$1.4   & 16.2$\pm$1.3   &  1.14$\pm$0.08 &  31.6$\pm$2.0 \\ 
LSB; IF 1  & 220.4 &  1.598$\pm$0.003    & 8.4$\pm$1.3   & 16.6$\pm$1.4   &  1.16$\pm$0.08 &  31.4$\pm$1.9  \\
LSB; IF 2  & 218.4 &  1.601$\pm$0.003    & 9.7$\pm$1.2   & 13.6$\pm$1.5   &  1.05$\pm$0.08 &  27.3$\pm$2.0  \\
\enddata
\tablecomments{I, Q, U fluxes, polarization fraction ($m$), and polarization position angles ($\chi$) at the four IF bands. Errors
in the I, Q, U fluxes are from image rms only.  They do not include the absolute flux calibration uncertainty which is the
same for all bands.}
      
\end{deluxetable}

\begin{figure}[ht] 
\begin{center} 
\vspace*{0.0 cm} 
\hspace*{0.0 cm} 
\includegraphics[angle=-90, scale=0.35]{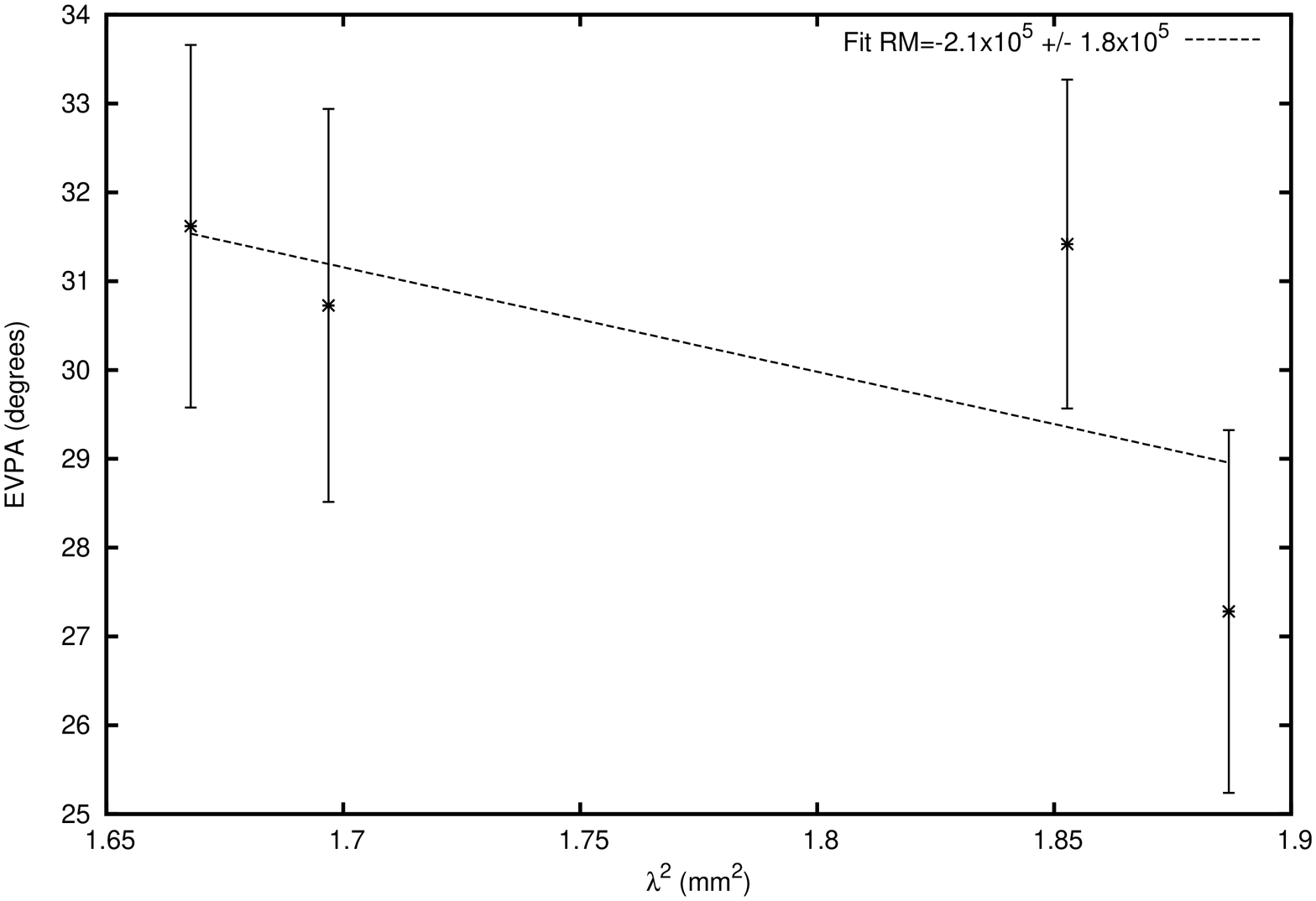}
\vspace*{0.0 cm} 
\caption{RM fit at the center of the M 87 core based on polarization position angles measured at four different frequencies. The best-fit RM is $-$(2.1$\pm$1.8)$\times$10$^{5}$ rad/m$^{2}$. The error bars are derived
from the 1$\sigma$ image uncertainty of the Q and U maps. }
\end{center} 
\end{figure}

In Figure 3 we show the EVPAs at the four frequencies with 1$\sigma$ error bars, together with a fit for the RM in the M 87 core. 
The best-fit RM is $-$(2.1$\pm$1.8)$\times$10$^{5}$ rad/m$^{2}$. 
Therefore, we determine the RM to be in the range between -7.5 $\times$ 10$^{5}$ and 3.3 $\times$ 10$^{5}$ rad/m$^{2}$ with a 3$\sigma$ confidence level. Since we are interested in an upper limit of the mass accretion rate, which is derived from the upper-limit (absolute value ) of RM (see section 4.1), we use 7.5 $\times$ 10$^{5}$ rad/m$^{2}$ for RM in the following discussion.

\section{Discussion}

\begin{figure}[ht] 
\begin{center} 
\vspace*{0.0 cm} 
\hspace*{0.0 cm} 
\includegraphics[angle=0, scale=0.4]{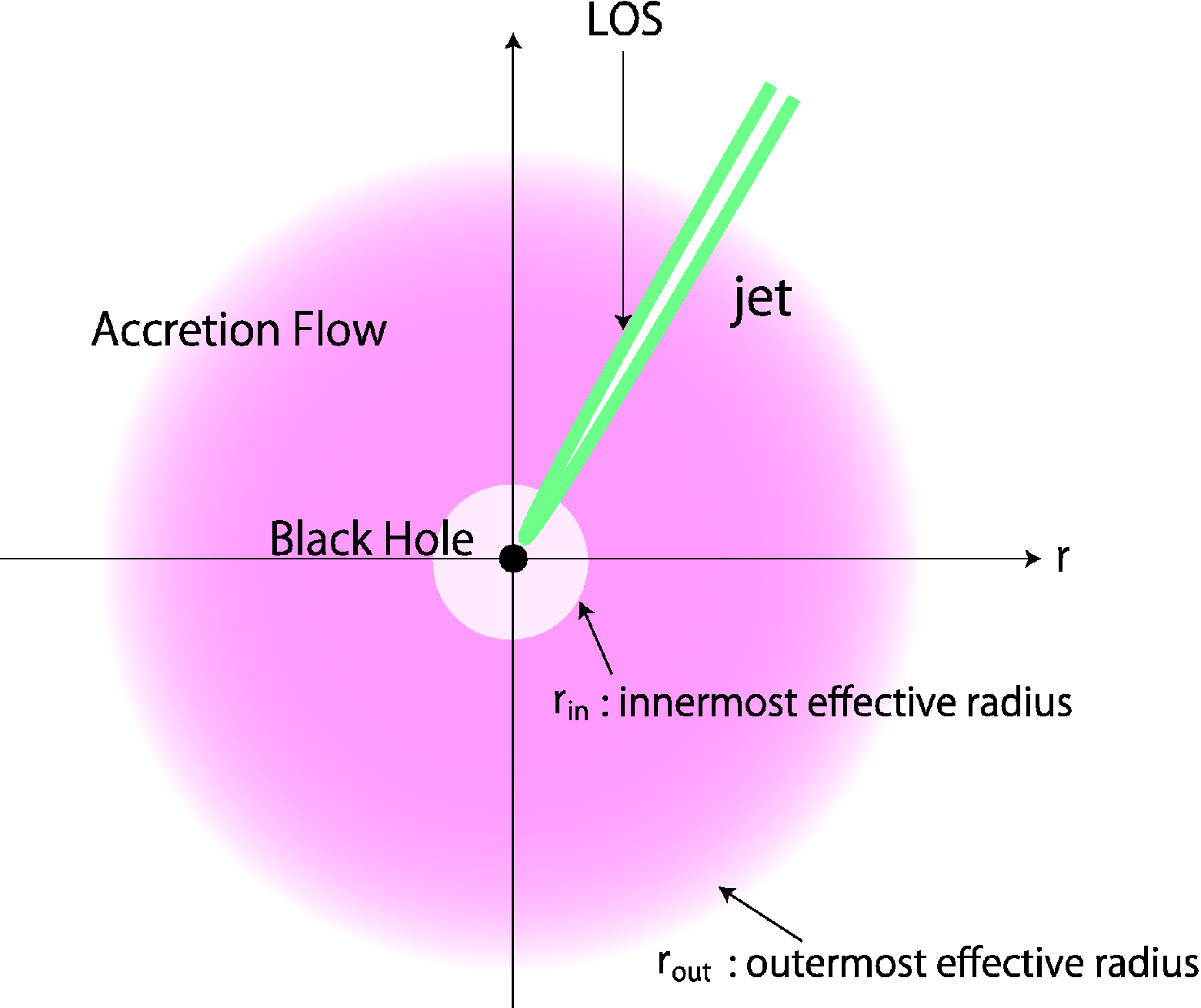}
\vspace*{0.0 cm} 
\caption{The envisioned accretion flow in M 87 for deriving the rotation measure. We assume that the M 87 jet offers the background polarized emission for 
Faraday rotation, and the dominant contribution comes from the innermost core of the jet. $r_{in}$ and $r_{out}$ are the inner and outer edge of the Faraday screen in the accretion flow where the electrons are sub-relativistic and the magnetic field is coherent.}
\end{center} 
\end{figure}  

\subsection{Constraints on Mass Accretion Rate and its Accretion Flow Model}

To constrain $\dot{M}$ with RM, we essentially follow the procedure outlined in Marrone et al. (2006).  
The primary assumption in this method is that the hot accretion flow in front of a source of synchrotron emission causes Faraday Rotation. The model also assumes that the density profile of the accretion flow follows a power-law distribution ($n$ $\propto$ $r^{-\beta}$) and that the magnetic field is well ordered, radial and of equipartition strength.  

In the case of M87, we further assume that the \emph{innermost} jet (i.e. its base) provides the dominant background polarized emission in the submillimeter band (see Figure 4). Note that while polarized emission could be emitted along the jet all the way from close to the BH to radii beyond $r_{\rm B}$, the dominant polarized emission at 230 GHz most likely originates from the jet base. This is because the majority of our SMA flux density (1.6 Jy) can be explained by the emission from the jet base at the center of M87. Based on the VLBI observation of M87 at 230 GHz, Doeleman et al. (2012) show that the jet base has a flux density of 1 Jy and a size of 5.5$r_{s}$. By using the core-shift relationship (Hada et al. 2011), Nakamura \& Asada (2013) determine the offset of the jet base from the BH at 230 GHz to be 4.3$r_{s}$ (in projected distance). Therefore, these studies imply that the dominant background emission quite likely originates from within a few $r_{s}$ from the BH. Nonetheless, we currently cannot exclude the possibility of polarized emissions from the diffuse part of the jet, and this issue will be addressed with future high-resolution observations.    
  
Based on Equation (9) in Marrone et al. (2006), we express $\dot{M}$ as a function of RM as 
\begin{equation*}
\dot M =2.2\times10^{-9}\left[1-(r_{out}/r_{in})^{-(3\beta-1)/2}\right]^{-2/3} \times \left( \frac{M_{\rm BH}}{6.6\times 10^9 M_{\odot}}\right)^{4/3} \left( \frac{2}{3\beta-1}\right)^{-2/3} r_{in}^{7/6} RM^{2/3} ~\rm rad/m^{2},
\end{equation*}
where $\beta$ is a parameter depending on the accretion flow models ($\beta$=1/2 $-$ 3/2 accounts for all varieties of accretion models). In the above equation, $r_{\rm in}$ and $r_{\rm out}$ are the inner and outer edge of the Faraday screen (in units of $r_{\rm s}$) where the electrons are sub-relativistic and the magnetic field is coherent. We remark that the electrons within $r_{\rm in}$ make a small or negligible contribution to RM because they become relativistic (i.e. $T_{e}>T_{\rm Rel}=6\times10^{9}$ K=$m_{e}c^{2}/k_{B}$) and RM is, thus, suppressed (Quataert \& Gruzinov 2000). 

In order to derive $\dot{M}$ with the above equation, it is important to determine $r_{\rm in}$ and $r_{\rm out}$. 
We estimate $r_{\rm in}$ based on the observed flux density of M87 from VLBI observations.  
As we described above, thermal electrons within $r_{\rm in}$ become relativistic and can emit significant synchrotron radiation that dominates the emission from the accretion disk (Yuan et al. 2003). Since it is expected that a hot accretion flow would be optically thick at cm wavelengths, 
the size of the region with relativistic plasma can be constrained by using the observed flux density and brightness temperature.  
For M 87, the observed total flux density of the nucleus is 1 Jy from the VLBI observation at 43 GHz (Abramowski et al. 2012).  
Since the brightness temperature of the hot electrons within $r_{\rm in}$ must be at least 6$\times$10$^{9}$ K, the diameter (i.e. 2~$r_{\rm in}$) of the emitting region of relativistic plasma can be constrained to be $\leq$ 0.32 mas (= 42$r_{\rm s}$). 
Since the observed flux density sets an upper limit on the amount of emission from relativistic electrons, it leads to an upper bound on $r_{\rm in}$ of 21$r_{\rm s}$. On the other hand, if the innermost jet resides right in front of the the emitting region of the relativistic plasma 
and the emission from the hot electrons is totally absorbed by the optically thick jet, 
we can not constrain the size of the region of the relativistic plasma directly from the observed flux. However, we can use the size of the innermost jet to constrain $r_{\rm in}$.  The observed size of the VLBI core at 43 GHz is 17 $\pm$ 4r$_{\rm s}$ (Asada \& Nakamura 2012).  
If the emission from the hot electrons is totally absorbed, $r_{\rm in}$ must be smaller than 17 $\pm$ 4r$_{\rm s}$.  
Therefore, the 21$r_{\rm s}$ upper bound for $r_{\rm in}$ is reasonable even for this case.  
We fix $r_{\rm out}$ to be at the Bondi radius $r_{\rm B}$. Here, $r_{\rm B}$ is just a convenient value to choose while $r_{\rm out}$ is not well determined. This is a reasonable choice because $\dot{M}$ is insensitive to $r_{\rm out}$, and $\dot{M}$ will only change by a factor of order unity even if $r_{\rm out}$ $<<$ $r_{\rm B}$. With these values, our new measurement of RM sets an upper limit $\dot{M}$ of 9.2 $\times$10$^{-4}$ M$_{\odot}$~yr$^{-1}$ at 21$r_{\rm s}$ for $\beta \leq 3/2$. 
This corresponds to 7.4 $\times$ 10$^{-3}$ $\dot{M}_{\rm B}$, which is at least two orders of magnitude smaller than the Bondi accretion rate. 
This value implies that the mass accretion rate is significantly suppressed while material is accreted and falling in from $r_{\rm B}$ to 21$r_{\rm s}$. 

Determining $\dot{M}$ provides an effective way to constrain accretion flow models because each model has its own unique $\dot{M}$ profile. 
The ADAF model requires $\dot{M}$ to be comparable to $\dot{M}_{\rm B}$ from $r_{\rm B}$ all the way to $r$$\sim$$r_{\rm s}$, 
while the ADIOS and the CDAF models require suppression of $\dot{M}$ as ($r/r_{\rm B}$)$^{\kappa}$$\dot{M}_{\rm B}$, 
where $r$ is the radius from the BH with 0 $<$ $\kappa$ $<$ 1 for ADIOS and $\kappa=1$ for CDAF.
Since our new results suggest significant suppression of $\dot{M}$ at the radius of 21$r_{\rm s}$, 
they disfavor the possibility of a classical ADAF and prefer ADIOS/CDAF.  
This result is in good agreement with the findings from nearly all numerical simulations (Yuan et al. 2012). 

The maximum radiative power that can be extracted from mass accretion is 
$P_{rad}$ $=$ $\eta \dot{M} / c^{2}$ $\simeq$ 7 $\times$10$^{45}$ $\eta$ $\dot{M} / \dot{M}_{\rm B}$ erg~s$^{-1}$, where $\eta$ is the radiative efficiency. With our upper limit of 9.2 $\times$10$^{-4}$ M$_{\odot}$~yr$^{-1}$ (= 7.4 $\times$ 10$^{-3}$ $\dot{M}_{\rm B}$), 
the maximum radiative power is estimated to be 5 $\times$ 10$^{42}$ erg~s$^{-1}$ if $\eta=0.1$.
This is about two orders of magnitude larger than the observed X-ray luminosity $L_{\rm X}$ of 7$\times$10$^{40}$ erg~s$^{-1}$, suggesting that the current limit of mass accretion rate is sufficient to explain the observed X-ray luminosity.  
On the other hand, the X-ray luminosity is low compared to the maximum available nuclear power. If $\dot{M}$ is not significantly lower than the upper limit obtained here, this will imply that the radiative efficiency is significantly lower than the canonical value of 0.1.  
In this case, both mass accretion rate and radiative efficiency are significantly suppressed which is consistent with the expected properties of RIAFs.

\subsection{Alternative interpretations ?}
Our upper limit on the rotation measure is derived from a set of polarization position angles (PA) with still significant errors. Because of the large uncertainties, it is difficult to demonstrate that PAs follow the $\lambda^{2}$ law, which is the basis of our fundamental assumption on the location of the Faraday screen (relative to the background source). As a result, we cannot completely rule out the scenario that $\emph{internal}$ Faraday rotation (i.e. Faraday screen intermixed with emitting plasma in the jet) plays an important role and the possibility that we are measuring an RM purely originating from cold electrons in the sheath of the jet, especially because the jet in M87 points to the observer nearly along the line of sight (i.e. $<$ 20$^{\circ}$). 

To assess the possibility of a jet-based RM, we adopt the Burn model (Burn 1966) on internal Faraday rotation and depolarization. By using the degree of linear polarization of the M87 core from the current study (1 \%) and that from the optical measurements (1\%-13\%; Perlman et al. 2011), we find that the maximum possible observed RM for the jet at 230 GHz is $\sim$6$\times$10$^{6}$ rad/m$^{2}$, beyond our current upper limit. Therefore, 
a jet-originated RM could provide an explanation for our observed RM. 

Nonetheless, in the context of our simplified one-dimensional model for spherically symmetric RIAFs, we still prefer the scenario that the Faraday rotating plasma (i.e. accreting gas) is external to the emission region because the dominant polarized emission quite likely comes from within a few $r_{s}$ from the BH (see section 4.1). In this case, since all the non-relativistic electrons reside beyond $\sim$21$r_{s}$ from the black hole, the Faraday screen must be external to the background source.  

In essence, in order to constrain $\dot{M}$ more unambiguously with the method used in this Letter, it is important to establish the $\lambda^{2}$ law of PA and obtain a more precise RM. This will rely on higher sensitivity observations with a significantly longer lever arm in the frequency space and higher angular resolution in the future. 

\acknowledgements
The SMA is a joint project between the Smithsonian Astrophysical Observatory and the Academia Sinica Institute of Astronomy and Astrophysics. We wish to thank all the staff members at the SMA who made these observations possible.

\end{document}